\documentclass[preprint,times]{elsarticle}
\usepackage{amsmath}
\usepackage{mathrsfs}
\usepackage{amsthm}
\usepackage{epstopdf}
\usepackage{graphicx}
\usepackage{xcolor}
\begin{document}
\begin{frontmatter}
\title{Modulation in degree of cross-polarization at Young's interferometer illuminated by non-uniformly polarized electromagnetic fields}
\author{ Rajneesh Joshi \textsuperscript{1,*}, and Gyaprasad \textsuperscript{2}} 
\address{\textsuperscript{1}Department of Physics, Government Degree College Danya, 263622 Almora, Uttarakhand, India
\\\textsuperscript{2}Department of Physics, Government Post Graduate College Khair, 202138 Aligarh, Uttar Pradesh, India

*Corresponding authors:- 
rajneeshjoshi077@gmail.com 
}
\begin{abstract}
The degree of cross-polarization (DoCP) and the electromagnetic degree of coherence (EM DoC) of an electromagnetic beam are investigated at the observation points for incoherent and non-uniformly polarized, i.e., different degree of polarization with respect to space at the two pinholes in Young's interference experiment. 
We note that DoCP at the observation plane can be expressed as the average value of the degree of polarization at the pinholes of Young's interference experiment. The present study also verifies that the DoCP is a two-point generalization of the DoP. Additionally, EM DoC at observation points can be controlled by DoP at pinholes. The findings may be useful for classical ghost imaging, lensless imaging, and the study of Stokes correlations.
\end{abstract}
\end{frontmatter}

\section{Introduction}
The statistical features of light fields, such as degree of coherence (DoC), degree of polarization (DoP), and degree of cross-polarization (DoCP) are mutually related and dependent quantities \cite{wolf2003unified, xin2008effect}. For electromagnetic (EM) fields, these quantities depend on the electric field correlations \cite{mandel1995optical}. DoC and DoCP are two-point correlation properties, whereas DoP is a one-point correlation property of light fields. Experimentally, these field correlations can be determined by visibility records using the interferometers suitable for measuring spatial and temporal coherence \cite{kanseri2013optical,kanseri2020determination,joshi2021relationship,turunen2022measurement}. The field's correlation at two-space points in the frequency domain is described by the cross-spectral density (CSD) matrix \cite{mandel1995optical}. For the single-point correlation, the CSD matrix is transformed into a spectral polarization matrix. Wolf's unified theory \cite{wolf2003unified} describes the possibility of measuring the change of polarization of a partially coherent EM beam during propagation. Later, for coherence and polarization determination, the generalized or coherence Stokes parameters are introduced \cite{korotkova2005generalized}. At single-point, the generalized Stokes parameters are reduced to the usual Stokes parameters. The spectral interference law for Stokes parameters describes that if two EM fields interfere, both intensity and polarization modulate \cite{setala2006contrasts,setala2006stokes}. The four contrast parameters (also known as the visibilities of modulations) are functions of intensity-normalized generalized Stokes parameters. Hence, the expressions of EM DoC, DoCP, and DoP can be written in terms of the intensity-normalized generalized Stokes parameters \cite{friberg2016electromagnetic}. Various techniques are available to measure the coherence and polarization in the literature \cite{kanseri2013optical,turunen2022measurement,friberg2016electromagnetic,kanseri2021measurement}.   

A laser is a fully spatially coherent source with a degree of coherence equal to 1. However, spatial coherence can also be generated from an incoherent source, such as an LED, through free-space propagation, as described by the van Cittert–Zernike theorem. The van Cittert-Zernike theorem connects the far-zone coherence and the polarization of the light source. In other words, the DoC of the radiated fields is obtained by the DoP of the light source and the Fourier transformation of the shape of the source \cite{tervo2013van}. In another study, it is shown that at certain pairs of points, the EM DoC of the radiated EM field depends on the DoP of the incoherent light source \cite{agarwal2005generation, kumar2025controlling}. Additionally, 
for a uniformly polarized beam, the value of DoCP at observation points equals the DoP of the incoherent source. Since the non-uniform polarization is the most general state of any light field, we aim to find the statistical properties at observation positions for these incoherent light beams. 
\begin{figure}[htbp]
\centering
\includegraphics[width=\linewidth]{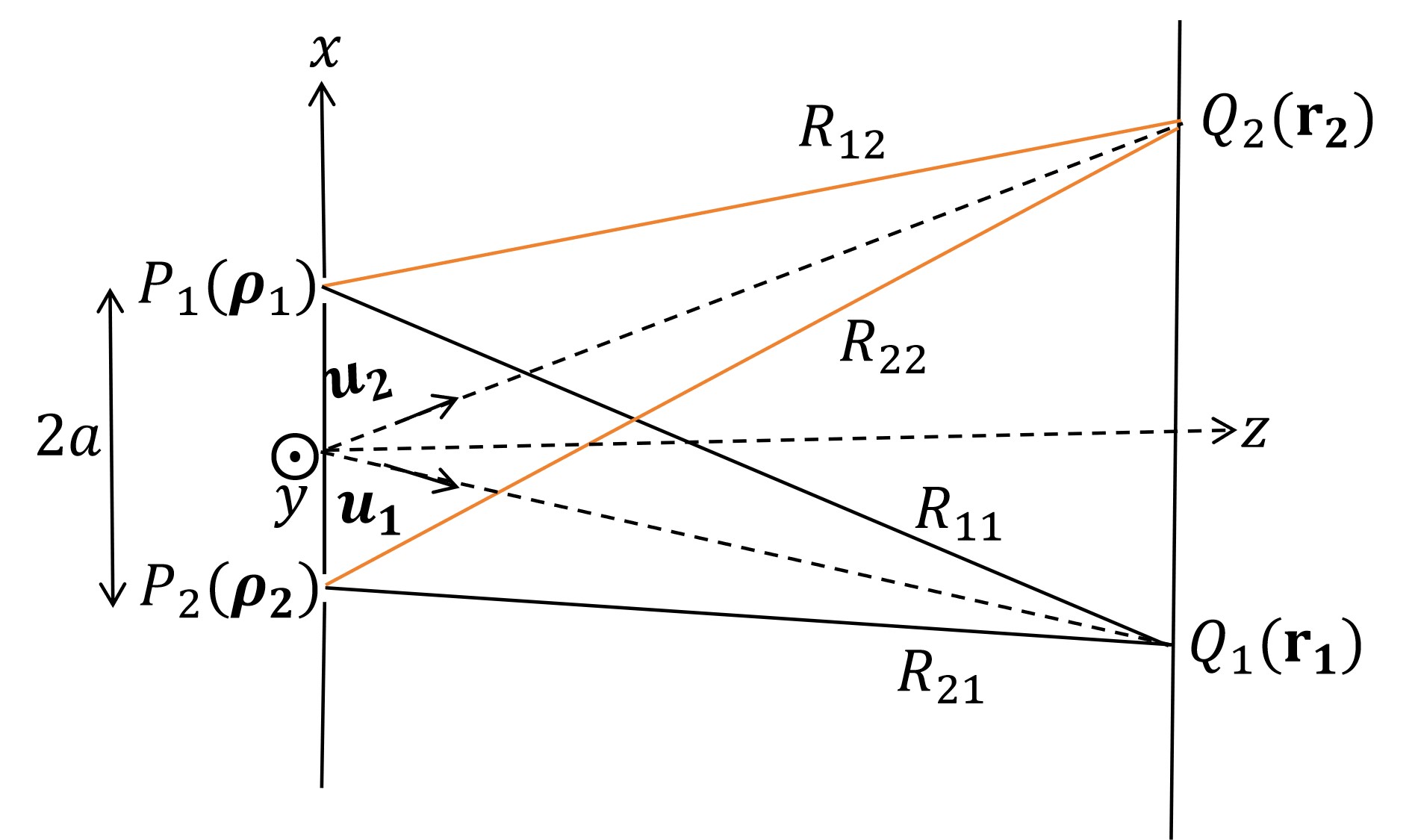}
\caption{Young's two-pinhole interference experiment to explore the statistical properties of random  electromagnetic fields at observation positions $Q_1(\textbf{r}_1)$ and $Q_2(\textbf{r}_2)$.}
\label{YIfig1}
\end{figure}

In this study, we first derive the expressions for the coherence Stokes parameters at the observation positions, which are expressed as combinations of the usual Stokes parameters and the coherence Stokes parameters at the pinholes of Young's interference experiment. 
Further, we calculate the DoCP and EM DoC at the observation points with the help of coherence Stokes parameters. We observe that both quantities are sinusoidally modulated for incoherent light with non-uniform polarization at the pinholes of Young's interference experiment. For a specific condition, both quantities depend only on the DoPs of the pinholes. For uniform polarization at the pinholes, the DoCP at the observation plane is equal to the DoP of the pinhole, as expected. 

\section{Origin and Development of the Degree of Cross-Polarization}
The correlations between the intensity fluctuations of light fields were determined by Hanbury Brown and Twiss (HBT) in 1950 \cite{padula2005hbt}. The HBT interferometer is used to measure the angular diameters of radio stars and has applications in condensed matter physics, high-energy physics, nuclear physics, and atomic physics \cite{shirai2007correlations}. Mathematically, when the correlation between intensity fluctuations is determined, a two-point quantity, called the degree of cross-polarization (DoCP), is obtained \cite{shirai2007correlations}. Examples of these intensity correlations can be observed in classical ghost imaging and lensless imaging. Initially, DoCP was derived in terms of the CSD matrix, which is assumed to be symmetric with respect to the two position vectors; i.e., the CSD matrix is Hermitian. For two identical points, it reduces to the usual DoP. For a scalar light field, such as a linearly $x$-polarized field, the DoCP is always 1. For unpolarized light with a factorized CSD matrix, in which the off-diagonal elements vanish and the diagonal elements are equal, the DoCP is zero \cite{shirai2007correlations}. Later, the DoCP was defined in terms of the coherence Stokes parameters \cite{volkov2008intensity}. It was also shown that two sources having different DoCP values, but the same spectral DoC and DoP can generate light beams with different DoPs \cite{xin2008effect}. In further development of DoCP research, the symmetry with respect to the position vectors in the CSD matrix was not assumed \cite{volkov2008intensity}, and an expression for the DoCP was derived, which is equal to the DoP for identical points. If a light beam is uniformly linearly polarized or unpolarized across its cross-section, the DoCP shares the same value as the DoP. However, if the light has orthogonal linear polarizations at two spatial positions, the DoCP becomes infinite \cite{volkov2008intensity}. The DoCP of EM beams has been studied in the cases of propagation through a turbulent atmosphere \cite{pu2009propagation}, free-space propagation \cite{sahin2009free}, and oceanic turbulence \cite{singh2025effect}  and its experimental validity has also been verified \cite{singh2022experimental}. Recently, the properties of the DoCP have been studied in the space-time domain \cite{kuebel2009properties}, and can be used to examine the relationship between polarization and coherence at two spatial points. For statistically stationary fields, if the absolute value of the DoC is unity, then $E(\boldsymbol{\rho}_2,t+\tau_0)=AE(\boldsymbol{\rho}_1,t)$, which implies that the field is statistically similar at the two points, where $\tau_0$ is the time delay. Kuebel  \cite{kuebel2009properties} showed that, under this condition, the DoCP between the two points is equal to the DoP at either of the two points. The converse of this theorem is generally not true. For a statistically stationary $x$-polarized beam, DoCP$ = $DoP$=1$, but the modulus value of the DoC is not necessarily equal to 1.  Kuebel also determined the DoCP for a statistically stationary electromagnetic field that is completely polarized at two points, i.e., $E_y(\boldsymbol{\rho}_1,t)=A_1E_x(\boldsymbol{\rho}_1,t)$, $E_y(\boldsymbol{\rho}_2,t+\tau_0)=A_2E_x(\boldsymbol{\rho}_2,t+\tau_0)$. In this case,  the modulus of the DoC is assumed to be non-zero. The DoCP depends on the amplitudes of the orthogonal polarization components and becomes unity when these amplitudes are equal. For fully coherent and fully polarized light, the DoCP shows the relationship between the states of polarization at two points on the Poincaré sphere, given by $\mathcal{P}(\boldsymbol{\rho}_1,\boldsymbol{\rho}_2,\tau)=\sqrt{\frac{3-\cos \theta}{1+\cos \theta}}$. Until now, the correlation between intensity fluctuations has been proportional to the spectral DoC defined in \cite{karczewski1963degree}. Later, the correlation between intensity fluctuations was derived in terms of the EM DoC defined by Tervo et al \cite{tervo2003degree, hassinen2011hanbury}. They showed that if linearly polarized light exists at two spatial positions that are perfectly correlated, then the EM DoC is 1. In this case, the DoCP is given by $\sqrt{2\sec^2 \theta -1}$, where $\theta$ denotes the orientation angle of the linear polarization at one point of the beam with respect to the horizontal polarization at the other point. The effect of scattering on the DoCP has also been studied \cite{jiang2019correlation}. Finally, the concept of the DoCP was introduced for nonstationary fields, such as pulsed light fields \cite{ding2011propagation}. For pulsed light fields, the coherence between two light frequencies is correlated and depends explicitly on the time instants as well as the time difference \cite{mandel1995optical}.

\section{Theoretical Background}
Now, we focus on the fundamental formulas related to the DoCP in terms of the normalized coherence Stokes parameters. We also present the definitions of the EM DoC, which are related to the DoCP.
The coherence property of a random EM beam is described by the CSD matrix. It is given by \cite{wolf2003unified}
\begin{align}
\label{eq1}
\boldsymbol{W}(\boldsymbol{\rho}_1, \boldsymbol{\rho}_2,\omega)=
\begin{bmatrix}
W_{xx}(\boldsymbol{\rho}_1, \boldsymbol{\rho}_2,\omega) & W_{xy}(\boldsymbol{\rho}_1, \boldsymbol{\rho}_2,\omega)\\
W_{yx}(\boldsymbol{\rho}_1, \boldsymbol{\rho}_2,\omega) & W_{yy}(\boldsymbol{\rho}_1, \boldsymbol{\rho}_2,\omega)
\end{bmatrix},
\end{align}
where the elements of above matrix are expressed by 
\begin{align}
\label{eq2} 
W_{jk}(\boldsymbol{\rho}_1, \boldsymbol{\rho}_2,\omega)=\langle E_j^*(\boldsymbol{\rho}_1,\omega)E_k(\boldsymbol{\rho}_2,\omega)\rangle, (j,k=x,y),
\end{align}
where the asterisk represents the complex conjugate, and the angular brackets represent the ensemble average.
The above matrix is very useful to find the correlation properties of a random EM beam and reduces to the spectral polarization matrix [$\textbf{W}(\boldsymbol{\rho},\boldsymbol{\rho},\omega)$] for equal spatial position, i.e., $\boldsymbol{\rho}_1=\boldsymbol{\rho}_2=\boldsymbol{\rho}$. The spectral polarization matrix and the usual Stokes parameters both define the polarization properties of random light fields. The connection between them is expressed by  \cite{mandel1995optical}
\begin{subequations}
\begin{align}
\label{eq3a}
S_0(\boldsymbol{\rho},\omega)=W_{xx}(\boldsymbol{\rho},\boldsymbol{\rho},\omega) +W_{yy}(\boldsymbol{\rho},\boldsymbol{\rho},\omega),   
\end{align}   
\begin{align}
\label{eq3b}
S_1(\boldsymbol{\rho},\omega)=W_{xx}(\boldsymbol{\rho},\boldsymbol{\rho},\omega) -W_{yy}(\boldsymbol{\rho},\boldsymbol{\rho},\omega),   \end{align}
\begin{align}
\label{eq3c}
S_2(\boldsymbol{\rho},\omega)=W_{yx}(\boldsymbol{\rho},\boldsymbol{\rho},\omega)+W_{xy}(\boldsymbol{\rho},\boldsymbol{\rho},\omega),\end{align}
\begin{align}
\label{eq3d}
S_3(\boldsymbol{\rho},\omega)=\iota[W_{yx}(\boldsymbol{\rho},\boldsymbol{\rho},\omega) -W_{xy}(\boldsymbol{\rho},\boldsymbol{\rho},\omega)].  \end{align}
\end{subequations}
The first and second Stokes parameters represent the summation and difference of spectral densities of horizontal and vertical electric field components, respectively. The last two Stokes parameters denote the difference between the spectral density of $45^\circ$, $135^\circ$, and right, left circular polarization, respectively. 
Later, the generalized Stokes parameters are also introduced, which determine the changes in the usual Stokes parameters during light propagation, and are defined as \cite{korotkova2005generalized}   
\begin{subequations}
\begin{align}
 \label{eq4a}
\mathcal{S}_0(\boldsymbol{\rho}_1,\boldsymbol{\rho}_2,\omega)=W_{xx}(\boldsymbol{\rho}_1,\boldsymbol{\rho}_2,\omega)+W_{yy}(\boldsymbol{\rho}_1,\boldsymbol{\rho}_2,\omega),   
\end{align}   
\begin{align}
 \label{eq4b}
\mathcal{S}_1(\boldsymbol{\rho}_1,\boldsymbol{\rho}_2,\omega)=W_{xx}(\boldsymbol{\rho}_1,\boldsymbol{\rho}_2,\omega)-W_{yy}(\boldsymbol{\rho}_1,\boldsymbol{\rho}_2,\omega),   
\end{align}
\begin{align}
\label{eq4c}
\mathcal{S}_2(\boldsymbol{\rho}_1,\boldsymbol{\rho}_2,\omega)=W_{yx}(\boldsymbol{\rho}_1,\boldsymbol{\rho}_2,\omega)+W_{xy}(\boldsymbol{\rho}_1,\boldsymbol{\rho}_2,\omega),
 \end{align}
\begin{align}
\label{eq4d} \mathcal{S}_3(\boldsymbol{\rho}_1,\boldsymbol{\rho}_2,\omega)=\iota[W_{yx}(\boldsymbol{\rho}_1,\boldsymbol{\rho}_2,\omega)-W_{xy}(\boldsymbol{\rho}_1,\boldsymbol{\rho}_2,\omega)],
  \end{align}
\end{subequations}
where $\mathcal{S}_0$ represents the sum of the correlation between horizontal and vertical electric field components, however $\mathcal{S}_1$, $\mathcal{S}_2$, and $\mathcal{S}_3$ correspond to the differences between the correlations for the $x$ and $y$, $45^{\circ}$ and $135^{\circ}$, and $r$ (right) and $l$ (left) circular polarizations, respectively, and $\iota$ represents the imaginary unit. For equal spatial positions, i.e., $\boldsymbol{\rho}_1=\boldsymbol{\rho}_2=\boldsymbol{\rho}$, the generalized Stokes parameters are reduced to usual Stokes parameters [$\mathcal{S}_n(\boldsymbol{\rho},\boldsymbol{\rho},\omega)=S_n(\boldsymbol{\rho},\omega)$] as shown in Eqs. (\ref{eq3a})-(\ref{eq3d}). Hence, the generalized Stokes parameters contain the information of coherence as well as polarization. The intensity-normalized generalized Stokes parameters, which are crucial parameters to define EM DoC and DoCP, are expressed by \cite{friberg2016electromagnetic} 
\begin{align}
\label{eq5}
\mu_n(\boldsymbol{\rho}_1, \boldsymbol{\rho}_2,\omega)=\frac{\mathcal{S}_n(\boldsymbol{\rho}_1, \boldsymbol{\rho}_2,\omega)}{\sqrt{S_0(\boldsymbol{\rho}_1, \omega)S_0( \boldsymbol{\rho}_2,\omega)}}, (n=0-3).    
\end{align}
The absolute value of intensity-normalized generalized Stokes parameters provides the modulation contrasts, $V_n(\omega)$.
Its relation to the EM DoC is given by \cite{tervo2003degree} 
\begin{align}
\label{eq6}   
\mu(\boldsymbol{\rho}_1, \boldsymbol{\rho}_2,\omega)
=\sqrt{\frac{1}{2}\sum_{n=0}^{n=3}|\mu_n(\boldsymbol{\rho}_1, \boldsymbol{\rho}_2,\omega)|^2}.
\end{align}
The EM DoC is a real quantity whose value lies in the range $0\leq \mu(\boldsymbol{\rho}_1, \boldsymbol{\rho}_2,\omega)\leq1$, where the limits 0 and 1 correspond to complete incoherence and complete coherence, respectively. It is important to note from Eqs. (\ref{eq5}) and (\ref{eq6}) that if there is no modulation in any of the Stokes parameters, i.e., $|\mu_n(\boldsymbol{\rho}_1,\boldsymbol{\rho}_2,\omega)|=0$, the light is incoherent. However, for fully coherent light, it is not possible for the modulation to reach the maximum value simultaneously in all Stokes parameters, i.e., $|\mu_n(\boldsymbol{\rho}_1,\boldsymbol{\rho}_2,\omega)|\neq 1$. 
The spectral DoC defined in \cite{karczewski1963degree} is the intensity-normalized zeroth coherence Stokes parameter, i.e., $\mu_0(\boldsymbol{\rho}_1,\boldsymbol{\rho}_2,\omega)$. In any light source, the intensity fluctuation is given by 
\begin{align}
\label{eq7}
\Delta I(\boldsymbol{\rho},\omega)= I(\boldsymbol{\rho},\omega)-\langle I(\boldsymbol{\rho},\omega) \rangle,   
\end{align}
where the first and second terms on the right-hand side denote instantaneous and averaged intensities, respectively.
 The correlation between intensity fluctuations is related to EM DoC as \cite{hassinen2011hanbury}
\begin{align}
\label{eq8}
\sqrt{\frac{\langle \Delta I (\boldsymbol{\rho}_1,\omega)\Delta I(\boldsymbol{\rho}_2,\omega)\rangle}{\langle I(\boldsymbol{\rho}_1,\omega)\rangle\langle I(\boldsymbol{\rho}_2,\omega)\rangle}}  =\mu(\boldsymbol{\rho}_1, (\boldsymbol{\rho}_2,\omega).  
\end{align}
After some straightforward calculations, DoCP can be written in terms of EM DoCs defined in \cite{karczewski1963degree} and \cite{tervo2003degree} as
\begin{align}
\label{eq9}
\mathcal{P} (\boldsymbol{\rho}_1,\boldsymbol{\rho}_2,\omega)=\sqrt{\frac{2\mu^2(\boldsymbol{\rho}_1,\boldsymbol{\rho}_2,\omega)}{|\mu_0(\boldsymbol{\rho}_1,\boldsymbol{\rho}_2,\omega)|^2}-1}.   
\end{align}
By substituting the value of EM DoC from Eq. (\ref{eq6}) into Eq. (\ref{eq9}), the DoCP is expressed in terms of normalized coherence Stokes parameters:
\begin{align}
\label{eq10} 
\mathcal{P}(\boldsymbol{\rho}_1,\boldsymbol{\rho}_2,\omega)=
|\mu_0(\boldsymbol{\rho}_1, \boldsymbol{\rho}_2,\omega)|^{-1}\sqrt{\sum_{n=1}^{n=3}|\mu_n(\boldsymbol{\rho}_1, \boldsymbol{\rho}_2,\omega)|^2}.
\end{align}
The DoCP is also a real-valued quantity and is bounded within the range $0\leq \mathcal{P}(\boldsymbol{\rho}_1,\boldsymbol{\rho}_2,\omega)\leq \infty$. It is a known fact that DoCP is a two-point property; hence, its value at two spatial positions is different for incoherent, fully coherent, and partially coherent light. 
At $\boldsymbol{\rho}_1=\boldsymbol{\rho}_2=\boldsymbol{\rho}$,  DoCP provides the value of DoP, which is connected to the intensity-normalized generalized Stokes parameters as \cite{al2007definitions}
\begin{align}
\label{eq11}  
P(\boldsymbol{\rho}, \omega)
=\frac{\sqrt{S_1^2(\boldsymbol{\rho}, \omega)+S_2^2(\boldsymbol{\rho}, \omega)+S_3^2(\boldsymbol{\rho}, \omega)}}{S_0(\boldsymbol{\rho}, \omega)}=\sum_{n=0}^{n=3}|\mu_n(\boldsymbol{\rho}, \boldsymbol{\rho},\omega)|^2-1.
\end{align}
The DoP takes values in the range $0\leq P(\boldsymbol{\rho}, \omega)\leq 1$, where the limits 0 and 1 correspond to unpolarized and completely polarized light, respectively, while intermediate values indicate partial polarization. DoP can be controlled in interferometric schemes \cite{leppanen2014interferometric, kanseri2018experimental}, using lens systems \cite{zhao2018controlling, sethuraj2020determination,sethuraj2021direct}, and also using electro-optically controlled liquid crystals \cite{kanseri2023degree,kanseri2024tunability,joshi2026tunable,gyaprasad2026tailoring,joshi2026real}. Also, an interferometric technique is discussed to tune both the degree of polarization and the degree of coherence \cite{kanseri2020development}.

\section{Theoretical aspects}  
We examine the behaviour of the DoCP and EM DoC at observation points that are produced from incoherent, nonuniformly polarized light in Young's two-pinhole experiment, as shown in Figure \ref{YIfig1}. In the paraxial approximation, the electric field components at observation points $\textbf{r}_1$ and $\textbf{r}_2$ can be expressed as \cite{agarwal2005generation}
\begin{align}
\label{eq12}  
E_j(\textbf{r}_1)=K\left[E_j(\boldsymbol{\rho}_1)\frac{\exp(\iota k R_{11})}{R_{11}}+E_j(\boldsymbol{\rho}_2)\frac{\exp(\iota k R_{21})}{R_{21}}\right],
\end{align}
\begin{align}
\label{eq13}  
E_j(\textbf{r}_2)=K\left[E_j(\boldsymbol{\rho}_1)\frac{\exp(\iota k R_{12})}{R_{12}}+E_j(\boldsymbol{\rho}_2)\frac{\exp(\iota k R_{22})}{R_{22}}\right],
\end{align}
where $(j=x,y)$, $K=\frac{-\iota \omega dA}{2\pi c}$, $\omega$ represents frequency, $dA$ is the area of pinholes, and $c$ is the speed of light. $R_{mn}, (m,n=1,2)$ denotes the distance from pinholes to observation positions, and can be written as 
\begin{align}
\label{eq14} 
R_{mn}\approx r_n-\boldsymbol{\rho}_m.\textbf{u}_n,
\end{align}   
where $r_n$ represents the modulus value of the position vector $\textbf{r}_n$ of observation points with respect to the origin. Hence, we can write as $\textbf{r}_1=r_1\textbf{u}_1$ and $\textbf{r}_2=r_2\textbf{u}_2$, where $\textbf{u}_1$ and $\textbf{u}_2$ are unit vectors along observation positions. $\boldsymbol{\rho}$ represents the position vectors of the pinholes of Young's interference experiment. It is expressed by $\boldsymbol{\rho}_1=a \hat{x}$, and $\boldsymbol{\rho}_2=a (-\hat{x})$. We also assume $r_1 \approx r_2=R$ in the further calculations of our study. 
By substituting Eqs. (\ref{eq12}), (\ref{eq13}), and (\ref{eq14}) into Eq. (\ref{eq2}), we obtain the elements of the CSD matrix as
\begin{align}
\label{eq15}  
W_{jk}(\textbf{r}_1,\textbf{r}_2)&=\frac{|K|^2}{R^2} \exp \left\{\iota k (r_2-r_1)\right\}\nonumber\\& \hspace{0mm}\times[W_{jk}(\boldsymbol{\rho}_1,\boldsymbol{\rho}_1)\exp\{\iota ka(u_{1x}-u_{2x})\}\nonumber \\ & \hspace{0mm}+W_{jk}(\boldsymbol{\rho}_2,\boldsymbol{\rho}_2)\exp\{-\iota ka(u_{1x}-u_{2x})\}\nonumber\\& \hspace{0mm}+W_{jk}(\boldsymbol{\rho}_1,\boldsymbol{\rho}_2)\exp\{\iota ka(u_{1x}+u_{2x})\}\nonumber \\ & \hspace{0mm}+W_{jk}(\boldsymbol{\rho}_2,\boldsymbol{\rho}_1)\exp\{-\iota ka(u_{1x}+u_{2x})\}]. 
\end{align}
\begin{figure}[htbp]
\centering
\includegraphics[width=0.8\linewidth]{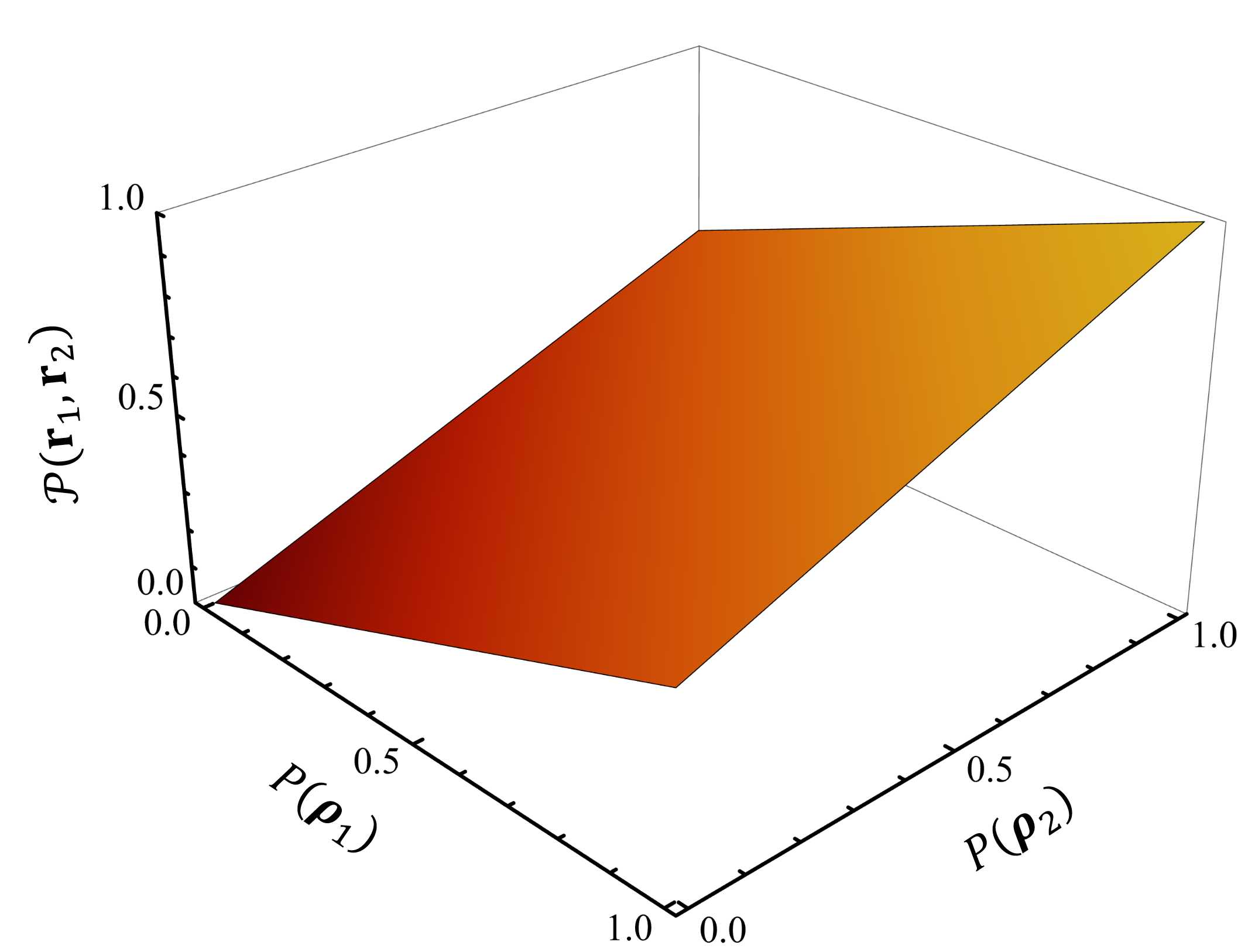}
\caption{Variation of the degree of cross-polarization at the observation points in Young's two-pinhole experiment with respect to the degrees of polarization at the pinholes.}
\label{Docpfig2}
\end{figure}
The above equation represents the CSD matrix at the observation plane and is ultimately related to the CSD matrix at the pinholes for all combinations of positions.
Further, the above equation can also be expressed in terms of coherence Stokes parameters as 
\begin{align}
\label{eq16}  
\mathcal{S}_n(\textbf{r}_1,\textbf{r}_2)&=\frac{|K|^2}{R^2} \exp \left\{\iota k (r_2-r_1)\right\}[S_n(\boldsymbol{\rho}_1)\exp\{\iota ka(u_{1x}-u_{2x})\}\nonumber \\ &\hspace{0mm}+S_n(\boldsymbol{\rho}_2)\exp\{-\iota ka(u_{1x}-u_{2x})\} \nonumber\\& \hspace{0mm}+\mathcal{S}_n(\boldsymbol{\rho}_1,\boldsymbol{\rho}_2)\exp\{\iota ka(u_{1x}+u_{2x})\}\nonumber \\ &\hspace{0mm}+\mathcal{S}_n(\boldsymbol{\rho}_2,\boldsymbol{\rho}_1)\exp\{-\iota ka(u_{1x}+u_{2x})\}]. 
\end{align}
The above equation represents the general form that describes the coherence Stokes parameters at observation points are proportional to the sum of coherence Stokes parameters of all input combinations.
By setting $\textbf{r}_2=\textbf{r}_1$ and $\textbf{r}_1=\textbf{r}_2$ in Eq. (\ref{eq16}), one obtains the usual stokes parameters at $\textbf{r}_1$ and $\textbf{r}_2$  as
\begin{subequations}
\begin{align}
\label{eq17a}
S_n(\textbf{r}_1)=\frac{|K|^2}{R^2} [S_n(\boldsymbol{\rho}_1)+S_n(\boldsymbol{\rho}_2) +\mathcal{S}_n(\boldsymbol{\rho}_1,\boldsymbol{\rho}_2)\exp\{2\iota kau_{1x}\} \hspace{0mm}+\mathcal{S}_n(\boldsymbol{\rho}_2,\boldsymbol{\rho}_1)\exp\{-2\iota kau_{1x}\}], 
\end{align}  
\begin{align}
\label{eq17b}  
S_n(\textbf{r}_2)=\frac{|K|^2}{R^2} [S_n(\boldsymbol{\rho}_1)+S_n(\boldsymbol{\rho}_2) +\mathcal{S}_n(\boldsymbol{\rho}_1,\boldsymbol{\rho}_2)\exp\{2\iota kau_{2x}\} \hspace{0mm}+\mathcal{S}_n(\boldsymbol{\rho}_2,\boldsymbol{\rho}_1)\exp\{-2\iota kau_{2x}\}]. 
\end{align}
\end{subequations}
\begin{figure}[htbp]
\centering
\includegraphics[width=0.8\linewidth]{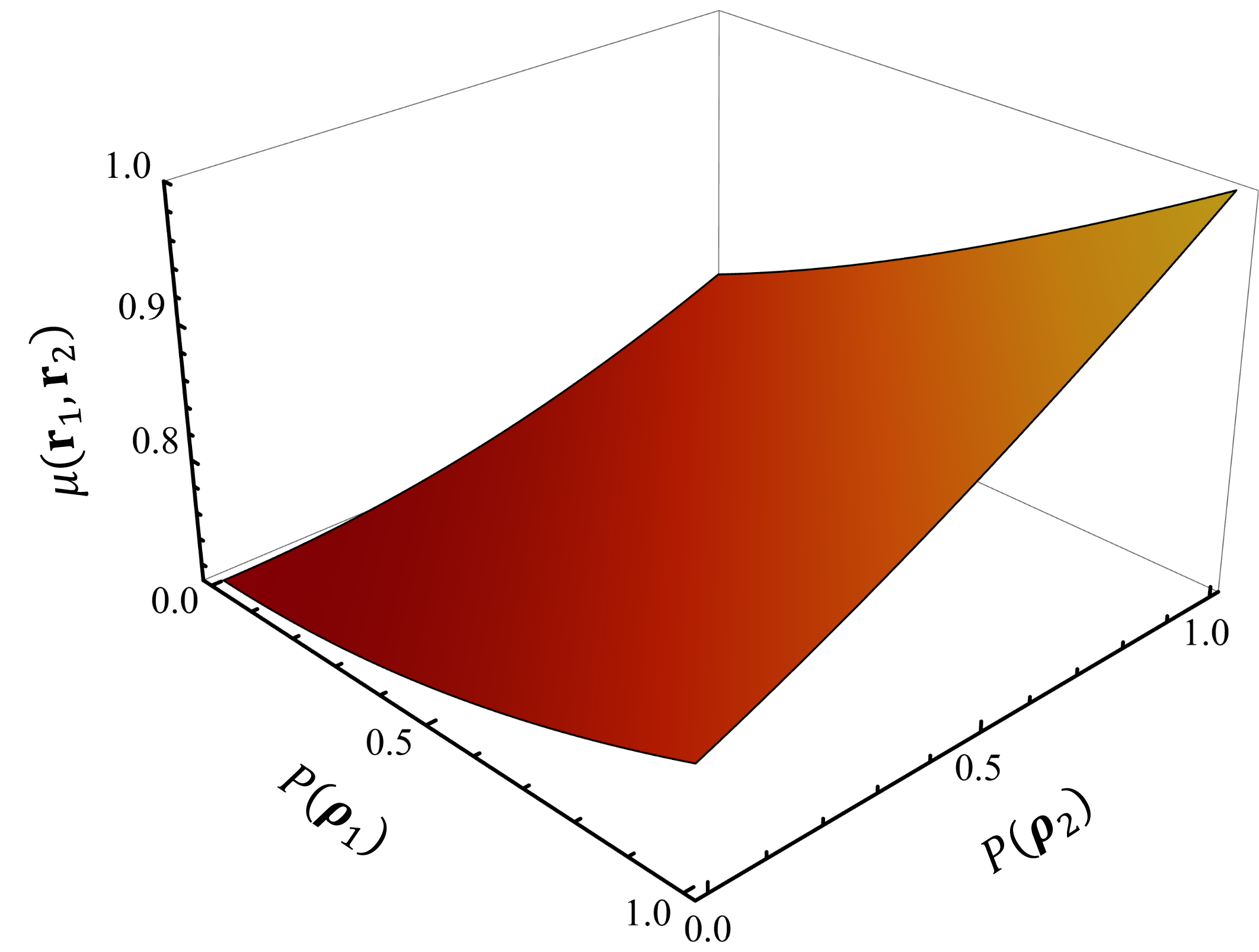}
\caption{Variation of the electromagnetic degree of coherence at two observation positions with respect to the degrees of polarization at the pinholes.}
\label{Emdocfig3}
\end{figure}
Assuming the light is incoherent at pinholes, the two-point correlation functions vanish. Substituting this condition into Eqs. (\ref{eq16}), (\ref{eq17a}), and (\ref{eq17b}) yields the expressions for DoCP at observation points
\begin{align}
\label{eq18}
\mathcal{P}^2(\textbf{r}_1,\textbf{r}_2)=\frac{P^2(\boldsymbol{\rho}_1)S_0^2(\boldsymbol{\rho}_1)+P^2(\boldsymbol{\rho}_2)S_0^2(\boldsymbol{\rho}_2)+2Re\{e^{2ika(\Delta u_x)}[\sum_{n=1}^{n=3}S_n(\boldsymbol{\rho}_1)S_n^*(\boldsymbol{\rho}_2)]\}}{S_0^2(\boldsymbol{\rho}_1)+S_0^2(\boldsymbol{\rho}_2)+2Re[e^{2ika(\Delta u_x)}S_0(\boldsymbol{\rho}_1)S_0^*(\boldsymbol{\rho}_2)]},   
\end{align}
where $\Delta u_x=u_{1x}-u_{2x}$, and $Re$ denote the real part. Equation (\ref{eq18}) clearly shows that DoCP is the function of non-uniform DoPs at pinholes. Additionally, it also depends on $\Delta u_x$ and Stokes vectors.
Similarly, the expression for EM DoC at observation points using Eqs. (\ref{eq16}), (\ref{eq17a}), and (\ref{eq17b}) is obtained as 
\begin{align}
\label{eq19}
\mu^2(\textbf{r}_1,\textbf{r}_2)=\frac{[1+P^2(\boldsymbol{\rho}_1)]S_0^2(\boldsymbol{\rho}_1)+[1+P^2(\boldsymbol{\rho}_2)]S_0^2(\boldsymbol{\rho}_2)+2Re\{e^{2ika(\Delta u_x)}[\sum_{n=0}^{n=3}S_n(\boldsymbol{\rho}_1)S_n^*(\boldsymbol{\rho}_2)]\}}{2[S_0^2(\boldsymbol{\rho}_1)+S_0^2(\boldsymbol{\rho}_2)+2S_0(\boldsymbol{\rho}_1)S_0(\boldsymbol{\rho}_2)]}.    
\end{align}
Equation (\ref{eq19}) shows that EM DoC is the function of DoPs at pinholes, $\Delta u_x$, and Stokes vector.
We note that the both DoCP and EM DoC sinusoidally modulate with $\Delta u_x$. 
For further calculation of DoCP and EM DoC, the Stokes vectors for the linearly partially polarized light can be expressed as \cite{kumar2011polarization}
\begin{align}
 \label{eq20}  
 \textbf{S}=[I, PI\cos2\theta, PI\sin2\theta, 0]^T,
\end{align}
where $\theta$ denotes the orientation of the linear polarization and $T$ denotes the transpose.
For the trivial case, $\Delta u_{x}=0$, and putting the values of Stokes vectors from Eq. (\ref{eq20}) into Eq. (\ref{eq18}),  expression for DoCP is obtained as 
\begin{align}
\label{eq21}
\mathcal{P}(\textbf{r}_1,\textbf{r}_2)=\frac{P(\boldsymbol{\rho}_1)+P(\boldsymbol{\rho}_2)}{2},    
\end{align}
and the EM DoC as 
\begin{align}
\label{eq22}
\mu(\textbf{r}_1,\textbf{r}_2)=\frac{1}{\sqrt{2}}\left[1+\frac{\{P(\boldsymbol{\rho}_1)+P(\boldsymbol{\rho}_2)\}^2}{4}\right]^\frac{1}{2}.    
\end{align}
Hence, the DoCP at an observation position  is the average value of the DoPs at pinholes. Figure \ref{Docpfig2} illustrates the variation of DoCP at observation plane as a function of DoP at two-pinholes. Three mutually perpendicular axis represent   $P(\boldsymbol{\rho}_1)$, $P(\boldsymbol{\rho}_2)$, and $\mathcal{P}(\textbf{r}_1,\textbf{r}_2)$, respectively. The color gradient from dark to light red depicts the transition from lower to higher DoCP values. It can be seen that the DoCP at observation plane is equal to the average  of DoPs at two pinholes. For example, if $P(\boldsymbol{\rho}_1)=0$ and $P(\boldsymbol{\rho}_2)=1$, the corresponding value of $\mathcal{P}(\textbf{r}_1,\textbf{r}_2)=0.5$. In the case of uniform polarization, where $P(\boldsymbol{\rho}_1)=P(\boldsymbol{\rho}_2)$, the DoCP becomes identical to DoP at each pinhole, i.e., $\mathcal{P}(\textbf{r}_1,\textbf{r}_2)=P(\boldsymbol{\rho}_1)=P(\boldsymbol{\rho}_2)$. Figure (\ref{Emdocfig3}) illustrates the variation of EM DoC, $\mu(\textbf{r}_1,\textbf{r}_2)$ as a function of DoP at pinholes. In this case, the three mutually perpendicular axes represent $P(\boldsymbol{\rho}_1)$, $P(\boldsymbol{\rho}_2)$, and $\mu(\textbf{r}_1,\textbf{r}_2)$, respectively. For example, when  $P(\boldsymbol{\rho}_1)=0$ and $P(\boldsymbol{\rho}_2)=1$, the EM DoC, $\mu(\textbf{r}_1,\textbf{r}_2)=0.79$. The minimum value of EM DoC is $\sqrt{\frac{1}{2}}$, which occurs when both pinholes contain completely unpolarized light, i.e., $P(\boldsymbol{\rho}_1)= P(\boldsymbol{\rho}_2)=0$. EM DoC reaches its maximum value of 1, when both pinhole contains fully polarized light.   

\section{Conclusion}
In conclusion, we derived the expression of coherence Stokes parameters, the degree of cross-polarization, and the electromagnetic degree of coherence at the observation positions of Young's interference experiment if light is incoherent but consists of non-uniform degree of polarization at its pinholes. Both the DoCP and EM DoC exhibit the sinusoidal behaviour with $\Delta u_x$. However, under the particular condition, i.e., $\Delta u_x=0$, both quantities depend solely on the degree of polarization at the pinholes. We found that the degree of cross-polarization at the observation points is the average value of the degree of polarization at both pinholes and becomes  equal to the degree of polarization only when the degree of polarization is uniform at both pinholes. This study demonstrates that the degree of cross-polarization is a generalized form of the degree of polarization. Furthermore, the electromagnetic degree of coherence varies with the DoPs at pinholes and attains its maximum value of unity only when the degree of polarization is 1 at each pinhole. The present findings provide insight into the characterization and manipulation of vector vortex beams, classical ghost imaging, and the measurement of intensity correlations between polarization components.   

\section*{Disclosures}
 The authors declare no conflicts of interest.
\bibliographystyle{elsarticle-num.bst}
\bibliography{references}

\end{document}